\documentclass[aps,pra,superscriptaddress,showpacs]{revtex4}

\usepackage{amsmath,amssymb,amsfonts}

\newcommand{\bra}[1]{\langle{#1}\vert}
\newcommand{\ket}[1]{\vert{#1}\rangle}

\begin{document}

\title{Operator formalism for the Wigner phase distribution}

\author{T. Subeesh}
\email{t.subeesh@gmail.com}
\affiliation{Dept. of Physics, Indian Institute of Technology Madras, Chennai 600036, India}
\affiliation{Dept. of Physics, Amrita School of Engineering, Coimbatore 641105, India}

\author{Vivishek Sudhir}
\email{vivishek.sudhir@gmail.com}
\affiliation{Dept. of Physics, Indian Institute of Technology Madras, Chennai 600036, India}
\affiliation{Dept. of Electrical \& Electronics, Amrita School of Engineering, Coimbatore 641105, India}

\begin{abstract}
The probability distribution for finding a state of the radiation field in a particular phase is described by a multitude of
theoretical formalisms; the phase-sensitivity of the Wigner quasi-probability distribution being one of them. We construct a
hermitian phase operator for this Wigner phase. We show that this operator is complete and also elucidate a set of complete but
non-orthogonal states that seems to be naturally associated with such an operator. Further we show that our operator satisfies
a weak equivalence relation with the Pegg-Barnett operator, thus showing that the essential phase information furnished by both
formalisms are the same. It is also shown that this operator gives results which are in correct agreement with the expected
uniform phase distribution of a Fock state.
\end{abstract}

%\keywords{Wigner phase operator}

\maketitle

\section{Introduction}

The problem of defining a Hilbert space operator for the quantum phase \cite{BarVac07} has been an outstanding one, a question that has plagued
physicists from the earliest days of quantum theory. The quantum phase is not only relevant to the radiation field, but also for
other bosonic systems, particularly Bose-Einstein condensates \cite{BarBurVac_BEC}. There has been significant progress in formulating a quantum phase
operator over the years \cite{BarVac07, nieto} , with the hermitian Pegg-Barnett phase formalism \cite{PegBar86, PegBar89} proving to be
very popular. The Pegg-Barnett formalism relies on states of well defined phase, given by,
\begin{equation} \label{phasestate}
\ket{\theta} = \lim_{s \rightarrow \infty} (s+1)^{-1/2} \sum_{n=0}^{s}\exp (i n \theta) \ket{n},
\end{equation}
which are defined in an $s-$dimensional Hilbert space and the limit $s \rightarrow \infty$ is understood to be taken
for physical expectation values; the associated phase operator is the projector,
\begin{equation}
    \rho_\text{PB}(\theta) = \ket{\theta}\bra{\theta}.
\end{equation}
Recently, there has been some debate over whether the Pegg-Barnett phase is the actual phase observed in experiments \cite{SkagBerg04}.

An alternative approach to defining a plausible quantum phase is, as the radial integral of one of the various
phase space quasi-probability distributions \cite{GarKni93, VargMoy04}. In this paper, we will focus on the phase probability
distribution defined as the radial integral over the Wigner distribution, viz.,
\begin{equation}\label{W_integral}
    P_{\psi}^\text{W}(\theta) = \int_0^\infty dr\, r\, W_\psi (r \cos \theta, r \sin \theta),
\end{equation}
where $W_\psi (x,p)$ is the Wigner transform of the state $\ket{\psi}$ and we refer to the above phase distribution as the Wigner phase
distribution. One of the main advantages of studying a phase probability defined in such a manner is the it becomes easier to understand
the appropriate classical limits via the Wigner-Weyl-Moyal correspondence. Moreover, Schleich \citep{Sch01} has suggested that
the phase states $\ket{\theta}$, defined in \eqref{phasestate}, can be represented as a wedge subtending an angle $\theta$ at the origin in phase space, suggesting
some relation between the Pegg-Barnett phase distribution and the Wigner phase distribution.

Our aim in this paper is to construct a hermitian operator for the Wigner phase distribution, to study its properties
and to elucidate its important Fock state and coherent state representations. Finally, we prove a weak-equivalence between the
Pegg-Barnett phase distribution and the phase distribution arising out of our Wigner phase operator. The final section of the
paper attempts to factorize the operator into a projector form.

\section{Wigner phase operator}

The phase sensitivity of various phase space distribution functions is the reason why they are a very attractive means of defining the quantum phase. In
particular, Schleich \citep{Sch01} has shown how to interpret this phase sensitivity in terms of phase space interference, which
is manifested as the ``overlap of areas'' in phase space. It is this same concept that prompted Knight \cite{GarKni92} to investigate the Wigner
phase \eqref{W_integral}.

We postulate that there exists a Wigner distribution associated with some phase state, such that this Wigner function is maximally sensitive to phase. This
means that this phase Wigner distribution, $W_\text{w}(r \cos \theta, r \sin \theta)$, is sharply localized around the phase angle $\theta$ i.e., if a state $\ket{\psi}$ has the Wigner distribution $W_\psi (x,p)$,
then $W_\text{w} (x,p)$ is characterized by the property,
\begin{equation}\label{Ww_delta_property}
\begin{split}
    \int_{0}^{2\pi} W_\psi (r \cos \theta', r \sin \theta') W_\text{w} (r \cos \theta' - \theta, r \sin \theta' - \theta) d\theta' =
        W_\psi (r \cos \theta, r \sin \theta).
\end{split}
\end{equation}
The Wigner phase operator that satisfies the above criteria, is given by the normal-ordered series,
\begin{equation}\label{rho_w}
\begin{split}
    \rho_\text{w} (\theta) = \frac{1}{2\pi}\, \sum_{m,n=0}^{\infty} \sum_{l=0}^n \frac{(-1)^m\; 2^{m+n/2}\; e^{i(n-2l)\theta}}{m!\; (n-l)!\; l!} \times\,
                        \Gamma\left( \frac{n}{2}+1 \right) \left( a^\dagger \right)^{m+n-l} a^{m+l},
\end{split}
\end{equation}
so that for a state represented by the density operator $\rho_\psi$, the phase probability distribution is given by,
\begin{equation*}
    P_\psi^\text{w} (\theta) = \text{Tr}\left[ \rho_\psi \rho_\text{w} (\theta) \right].
\end{equation*}
The above expression \eqref{rho_w} is derived by using the property \eqref{Ww_delta_property} on the Wigner function of the coherent state
$\ket{\alpha} = D(\alpha)\ket{0}$ and then using the correspondence \citep{CahGlau69} between normal ordered operator expressions and the coherent
state expectations of the operator.

\subsection{Representations of the operator}

The form of the Wigner phase operator, in its normal-ordered form as given in \eqref{rho_w}, is not tenable for easy manipulation and is in general quite complicated.
Any application of this operator formalism for actual computation would necessitate an understanding of the associated representations in various basis. Of course,
the two most important bases used frequently in the study of the radiation field is the Fock basis and the coherent state basis.

It can be shown, after some algebraic gymnastics, that the number state representation of $\rho_\text{w}(\theta)$ is given by,
\begin{eqnarray}\label{rhow_sr}
    \bra{s} \rho_\text{w}(\theta) \ket{r} &=& \sqrt{\frac{s!\, r!}{4\pi^2}} \times
        \begin{cases} \label{rhow_sr1}
            \sum_{n=0}^s \frac{(-1)^{s-n} (\sqrt{2}e^{-i\theta})^{r-s+n} (\sqrt{2}e^{i\theta})^{n} }{n!(s-n)!(r-s+n)!}
                \Gamma \left( \frac{r-s}{2}+ n+ 1 \right) & \text{if $r \geq s$} \\
            \sum_{n=0}^r \frac{(-1)^{r-n} (\sqrt{2}e^{-i\theta})^{n} (\sqrt{2}e^{i\theta})^{s-r+n} }{n!(r-n)!(s-r+n)!}
                \Gamma \left( \frac{s-r}{2}+ n+ 1 \right) & \text{if $s \geq r$}
        \end{cases} \\
    &=& \sqrt{\frac{s!\,r!}{4\pi}} \times
        \begin{cases} \label{rhow_sr2}
            \frac{(-1)^s}{s!}\, \left(\frac{e^{-i\theta}}{\sqrt{2}}\right)^{r-s}\,
                \frac{{}_2 F_1 \left( -s, 1+ \frac{r-s}{2} ; 1+r-s ; 2 \right)}{\Gamma \left( \frac{1}{2}+\frac{r-s}{2} \right)} & \text{if $r \geq s$} \\
            \frac{(-1)^r}{r!}\, \left(\frac{e^{i\theta}}{\sqrt{2}}\right)^{s-r}\,
                \frac{{}_2 F_1 \left( 1+ \frac{s-r}{2}, -r ; 1+s-r ; 2 \right)}{\Gamma \left( \frac{1}{2}+\frac{s-r}{2} \right)} & \text{if $s \geq r$}.
        \end{cases}
\end{eqnarray}
This expression clearly shows the stark departure of the Wigner phase operator from the Pegg-Barnett phase operator; nevertheless, for the important case $s=r$, both
the formalisms give a uniform phase probability distribution, $P^\text{w}(\theta) = P^\text{PB}(\theta) = \frac{1}{2\pi}$. Moreover, it is easy to see from \eqref{rhow_sr1}
that $\rho_\text{w}(\theta)$ is a hermitian phase operator.

Given two coherent states $\ket{\alpha}$ and $\ket{\beta}$, the Wigner phase operator can be represented in the Bargmann basis as,
\begin{eqnarray}\label{rhow_alpha_beta}
    \bra{\beta}\rho_\text{w}(\theta)\ket{\alpha} = \frac{\langle \beta \ket{\alpha} e^{-2\alpha \beta^*}}{2\pi}\, \left[ 1+ \sqrt{\pi}\,z e^{z^2}
        \left( 1+ \text{erf}(z) \right) \right],
\end{eqnarray}
where $z=\frac{1}{\sqrt{2}}( \alpha e^{-i\theta} + \beta^* e^{i\theta} )$ and $\text{erf}(z)$ stands for the error-function of the argument $z$. For the case
$\beta= \alpha$, the above expression leads to the phase probability distribution of the coherent state $\ket{\alpha}$, given by,
\begin{equation}\label{rhow_alpha_alpha}
    P^\text{w}_{\alpha}(\theta) = \frac{e^{-2\vert \alpha \vert^2}}{\pi} \left[\frac{1}{2}+\sqrt{\frac{\pi}{2}}\, a e^{2a^2} \left( 1+ \text{erf}(a\sqrt{2})
    \right) \right],
\end{equation}
where, $a = \vert \alpha \vert \cos (\theta - \text{arg}\, \alpha)$. This is the same expression that was derived by Knight \citep{GarKni92, GarKni93} using the radial
integral \eqref{W_integral}. From the properties of the error-function, it is trivial to note that the above Wigner phase distribution for the coherent state is a
positive quantity, thus enabling it to be interpreted as a true probability distribution.

An important piece of geometrical insight can be obtained from a slight manipulation on the above expression for the Wigner phase distribution of the coherent state. It
is possible to prove, starting from \eqref{rhow_alpha_alpha}, that,
\begin{equation*}
    \int_0^{2\pi} d\theta\; e^{im\theta}\; \bra{\alpha} \rho_\text{w}(\theta)\ket{\alpha} =
        e^{im\; \text{arg}\; \alpha}\; e^{-2\vert \alpha \vert^2}\; \left(\frac{\vert \alpha \vert}{\sqrt{2}}\right)^m
        \frac{\sqrt{\pi}}{\Gamma \left( \frac{m+1}{2} \right)}\; {}_1 F_1 \left( \frac{m}{2}+1; m+1; 2\vert \alpha \vert^2 \right),
\end{equation*}
for any integer $m$. Now taking the limit $\vert \alpha \vert \rightarrow \infty$ on either side, we find,
\begin{equation*}
    \lim_{\vert \alpha \vert \rightarrow \infty} \int_0^{2\pi} d\theta\; e^{im\theta}\; \bra{\alpha} \rho_\text{w}(\theta)\ket{\alpha} = e^{im\; \text{arg}\; \alpha},
\end{equation*}
which implies that,
\begin{equation}\label{lim_alpha_infty}
    \lim_{\vert \alpha \vert \rightarrow \infty}\; \bra{\alpha} \rho_\text{w}(\theta)\ket{\alpha} = \delta (\theta - \text{arg}\; \alpha).
\end{equation}
Here, $\delta(\theta - \theta')$ is the angular delta function characterized by the property $\int_0^{2\pi} f(\theta') \delta(\theta - \theta') d\theta' = f(\theta)$.

The way to interpret \eqref{lim_alpha_infty} is to first note that the Wigner function of the state $D(\alpha) \ket{\psi}$ is the shape-preserved displacement of the
Wigner function of $\ket{\psi}$, with the displacement affected in the $\text{arg}\; \alpha$ direction by a distance proportional to $\vert \alpha \vert$. It is also
well known that the phase distribution of the coherent state $\ket{\alpha}$ is peaked about $\text{arg}\; \alpha$. Now, \eqref{lim_alpha_infty} is a mathematical
statement of the fact that as the ``hump'' shaped coherent state Wigner function is displaced further away from the origin, the ``hump'' completely falls within the
angular wedge that subtends an angle $\theta$ at the origin, resulting in a sharply defined phase. This geometric interpretation carried by the Wigner phase operator is
inherited from the properties of the radially integrated Wigner phase distribution, the latter of which has been noted previously \citep{Sch01}. From this geometric
idea associated with the Wigner phase operator, it becomes evident that the number state, whose Wigner function is centered at the origin, should have a uniform phase
probability distribution, as has been mathematically proved above. Further, any state whose Wigner function can be localized within an angular wedge is expected to have
a correspondingly localized phase probability distribution.

\subsection{Properties of the operator}

We have already shown in the previous section, using the Fock state representation \eqref{rhow_sr}, that $\rho_\text{w}^\dagger (\theta) = \rho_\text{w}(\theta)$ i.e.,
that the Wigner phase operator is hermitian. In this section, we will enumerate some more of the important properties of the Wigner phase operator.

\subsubsection{Completeness}

Together with hermiticity, completeness is an essential property of an infinite dimensional operator which ensures that its orthonormal eigenstates span
the associated Hilbert space. Before we show that $\rho_\text{w}(\theta)$ is complete, we make use of a Fock decomposition of the operator \eqref{rho_w}, by inserting a set of complete Fock states between the raising and lowering operator powers. Performing the subsequent algebraic simplifications, we get the expression,
\begin{equation*}\label{rhow_fock_decomp}\begin{split}
\rho_\text{w}(\theta) = \frac{1}{2\pi} \sum_{n,k=0}^{\infty} \sum_{l=0}^n \Gamma\left(\frac{n}{2}+1\right) \frac{(-1)^k}{k!}
                    \frac{\left(e^{-i\theta}\sqrt{2}\right)^{n-l}}{(n-l)!} \frac{\left(e^{i \theta}\sqrt{2}\right)^l}{l!} \times \\
                    \sqrt{(k+l)!(n+k-l)!}\; \ket{k+l} \bra{n+k-l}.
\end{split}\end{equation*}
Using the above form of the operator, we can perform a straightforward integration of either sides to obtain the completeness result,
\begin{equation}\label{rhow_complete}
   \int_0^{2\pi} \rho_\text{w}(\theta)\, d\theta = 1.
\end{equation}

From the geometric interpretation associated with the Wigner phase operator expounded in the previous section, it becomes clear that the above completeness property is
intimately related to the requirement that the Wigner function be normalized.

\subsubsection{Relation with Pegg-Barnett phase operator}

Since the state $\ket{\theta}$ defined in \eqref{phasestate} are states of well-defined phased, it is of interest to investigate the phase of these states as predicted by
our operator formalism. The Wigner function of these phase states has been worked out previously \citep{Herz93}. Here, we attempt to evaluate the Wigner phase distribution
$P^\text{w}_{\phi}(\theta)$ associated with the phase state $\ket{\phi}$. Before we do this, we give a Fock state decomposition of $\rho_\text{w}(\theta)$ which is very
conducive to subsequent developments, viz.,
\begin{equation}\label{rhow_outer_prod1}
    \rho_\text{w}(\theta) = \frac{1}{2\pi} \sum_{n,k=0}^\infty \Gamma \left( \frac{n+k}{2}+1 \right) \frac{(\sqrt{2}e^{i\theta})^k}{k!} (a^\dagger)^k
            \left[ \sum_{m=0}^\infty (-1)^m \ket{m}\bra{m} \right] a^n \frac{(\sqrt{2}e^{-i\theta})^n}{n!}.
\end{equation}
Using the above expression to evaluate $P^\text{w}_{\phi}(\theta)$ we get the following elegant result,
\begin{equation}\label{Pw_PB}
    P^\text{w}_{\phi}(\theta) = \text{Tr}\left[ \rho_\text{w}(\theta -\phi) \rho_\text{PB}(0) \right],
\end{equation}
which implies that whenever four angles satisfy the relation, $\theta - \phi = \theta' - \phi'$, we have the following relation,
\begin{equation}\label{weak_equiv}
    \text{Tr}\left[ \rho_\text{w}(\theta) \rho_\text{PB}(\phi) \right] = \text{Tr}\left[ \rho_\text{w}(\theta') \rho_\text{PB}(\phi') \right].
\end{equation}
We term the above relation, the weak-equivalence between the Pegg-Barnett and the Wigner phase operator formalisms. \eqref{Pw_PB} is a statement of the fact that when the
phase of a state is well-defined, then its phase, as given by its expectation over $\rho_\text{w}(\theta)$, gives the phase only upto a relative reference phase; this is of
course a very important property for any phase operator. Similarly, another way of looking at the weak-equivalence \eqref{weak_equiv} is to note that either side of the
equation is actually the Hilbert-Schmidt inner product of the respective operator pairs. Thus, in some sense, the equation claims that the ``overlap'' between the
Wigner and Pegg-Barnett phase operators is invariant under the transformation $\theta - \phi = \theta' - \phi'$.

\subsection{Operator in projector form}

A very important aspect of any hermitian operator is its projection property i.e., to identify the space onto which a measurement of the associated physical property
would finally leave the system in. Whenever an operator has a diagonal representation in terms of some orthogonal basis, it is a projector when it is
idempotent i.e., when the operator squares to itself. The Wigner phase operator is definitely not idempotent, as can be easily deduced from \eqref{rhow_outer_prod1}.

We use the integral representation of the gamma function, viz., $\Gamma(x) = 2 \int_0^\infty u^{2x-1}e^{-u^2} du$ in \eqref{rhow_outer_prod1}, to express it in the form,
\begin{equation}\label{rhow_outer_prod2}
    \rho_\text{w}(\theta) = \frac{1}{\pi} \int_0^\infty du\; u e^{-u^2} \sum_{m=0}^{\infty} (-1)^m \ket{\sqrt{2}u e^{i\theta},m}\bra{\sqrt{2}u e^{i\theta},m},
\end{equation}
where the state $\ket{\sqrt{2}u e^{i\theta},m}$ is a particular case of the general state,
\begin{equation}\label{ket_z_m}
    \ket{z,m} = \sum_{n=0}^\infty \frac{z^n}{n!} (a^\dagger)^n \ket{m}.
\end{equation}
Together with the above definition, \eqref{rhow_outer_prod2} provides a diagonal representation of the Wigner phase operator, in the sense that the integral and sum
contains only terms that couple projectors of the form $\ket{z,m}\bra{z',m'}$ for $z=z', m=m'$. Note that the state $\ket{z,m}$ is not normalized as given in
\eqref{ket_z_m}; to normalize it, it has to be divided by a factor of $\sqrt{I_0 (2 \vert z \vert)}$, where $I_\nu$ is the Bessel function of the second kind of order
$\nu$. We also have the relation for the un-normalized states,
\begin{equation*}
    \bra{z',m'}z,m \rangle = \frac{z^m ({z'}^*)^{m+m'}}{(m+m')!}\sqrt{\frac{m!}{m'!}}\, {}_1 F_1 \left( m+1; m+m'+1; {z'}^* \right),
\end{equation*}
thus showing that these states do not furnish an orthogonal set. Owing to the completeness relation \eqref{rhow_complete}, we have the resolution of unity,
\begin{equation*}
    \int dz\; dz^*\; e^{-\vert z \vert^2} \sum_{m=0}^\infty (-1)^m \ket{z,m}\bra{z,m} = 1,
\end{equation*}
so that these states do furnish a complete bases set, although not orthogonal, very much like the Glauber coherent states.

\section{Conclusions}

The hermitian Wigner phase operator that we have constructed is shown to give the expected uniform phase distribution for a number state, while for a coherent state, it
gives a distribution which essentially captures the same phase information as given by the Pegg-Barnett formalism. Subsequently, it is shown that our operator is
complete and we also give a diagonal representation of the same in terms of a set of complete states which seem to be naturally suited for describing phase in the quantum
phase space. Finally, we also prove that the Wigner phase operator satisfies a weak-equivalence relation with the Pegg-Barnett operator, which is the reason why the radially integrated Wigner function captures essentially the same phase information as the latter.

The concept of defining a phase via the Wigner function gives a more general way to understand phase and it maybe the case that one can ascribe a consistent meaning to the
phase of non-bosonic systems and the authors are pursuing work along these directions.

\begin{acknowledgments}
One of the authors (TS) would like to thank Prof. P. C. Deshmukh at the Indian Institute of Technology Madras for a valuable suggestion
on the nature of infinite dimensional operators, made during the course of the work. The other author would like to thank Prof. N. Mukunda at the Indian Institute of
Science for helpful discussions regarding the notion of a Wigner distribution and the associated generalizations.
\end{acknowledgments}

\bibliography{ref_wigner_phase}

\end{document}